\documentclass{ar-1col-S2O}
\usepackage{url}
\usepackage[normalem]{ulem}
\usepackage{lineno}
\usepackage{amsmath,amssymb}
\newcommand{\snn}{\ensuremath{ \sqrt{s_{\rm NN} } } }

\usepackage[
  backend=biber,
  style=numeric-comp,
  giveninits=true,
  defernumbers=true,
  doi=false,
  sorting=none
]{biblatex}
\DeclareFieldFormat{pages}{#1}
\renewbibmacro*{in:}{}
\AtEveryBibitem{
\clearfield{title}
\clearfield{booktitle}
\clearfield{maintitle}
}

\DeclareSourcemap{
  \maps[datatype=bibtex]{
    \map{
      \step[fieldsource=collaboration, final]
      \step[fieldset=author, origfieldval]
      \step[fieldset=author, append, fieldvalue={ Collaboration}]
    }
  }
}

\DeclareSourcemap{
  \maps[datatype=bibtex]{
    \map{
      \step[fieldsource=collaboration, final]
      \step[fieldset=usera, origfieldval]
      \step[fieldset=usera, append, fieldvalue={ Collaboration}]
    }
  }
}

\renewbibmacro*{author}{%
  \ifboolexpr{ test \ifuseauthor and not test {\ifnameundef{author}} }
    {\printnames{author}}
    {\global\undef\bbx@lasthash}%
  \iffieldundef{usera}
    {}
    {\setunit{\addspace}\printtext{(\printfield{usera})}}%
}

\jname{Xxxx. Xxx. Xxx. Xxx.}
\jvol{AA}
\jyear{YYYY}
\doi{annurev-nucl-100324-014902 (will become active upon publication)}

\begin{document}

\markboth{P. Braun-Munzinger, A. Rustamov, N. Xu}{The phase structure of
QCD}


\title{The phase structure of QCD: Fluctuations and Correlations}

\author{P. Braun-Munzinger,$^{1,2,3}$ A. Rustamov,$^{4, 5}$ \\ and Nu Xu $^{3,6}$
\affil{$^1$Extreme Matter Institute EMMI, GSI Helmholtzzentrum f{\"u}r Schwerionenforschung, Darmstadt, Germany, 64291 }
\affil{$^2$Physikalisches Institut, Universit\"{a}t Heidelberg, Heidelberg, Germany, 69120}
\affil{$^3$Key Laboratory of Quark and Lepton Physics (MOE), Central China Normal University, Wuhan, China, 430079}
\affil{$^4$ GSI Helmholtzzentrum f{\"u}r Schwerionenforschung, Darmstadt, Germany, 64291}
\affil{$^5$National Nuclear Research Center, Baku, Azerbaijan, 1069}
\affil{$^6$Institute of Modern Physics, Chinese Academy of Sciences, Lanzhou, China, 730000}}

\begin{abstract}
The strong interaction - governed by Quantum Chromodynamics (QCD) - shapes the structure of the visible universe. At about 10 $\mu$s after the big bang, the primordial matter made up of quarks and gluons plus leptons, photons and neutrinos, the quark-gluon plasma (QGP), became cool enough to create, in a phase transition, the protons and neutrons of ordinary matter, along with other strongly interacting unstable hadrons. This phase transition was predicted within the framework of QCD and has been studied in accelerator laboratories world-wide since about 40 years. This review will explore recent breakthroughs in the study of the QCD phase diagram. We will highlight measurements of particle production and fluctuations, and compare them to theoretical predictions. We summarize our current understanding of the QCD structure and outline future experimental opportunities with high energy nuclear collisions at fixed-target and collider facilities world-wide.
\end{abstract}


\maketitle

\tableofcontents

\section{Introduction} \label{int}
The matter formed in relativistic nuclear collisions is governed by Quantum Chromodynamics (QCD), the theory of the strong interactions. Over the past 40 years, experiments at the BNL AGS and RHIC and the CERN SPS and LHC accelerators have established that, in such collisions, strongly interacting matter is formed with properties close to that of an ideal fluid~\cite{Busza:2018rrf}. Importantly, when hadrons, the visible constituents of the world we live in, are formed, the state of the matter is very close to  thermal equilibrium in such collisions. For reviews see ~\cite{Gyulassy:2004zy,Braun-Munzinger:2015hba, Andronic:2017pug,Bzdak:2019pkr,ALICE:2022wpn}. 

The subject of this review is the phase structure of the matter created in high-energy nuclear collisions. Progress in delineating this phase structure has been obtained within the framework of 'lattice QCD', i.e., by numerically solving the equations of QCD on a discrete space-time lattice and extrapolating to vanishing lattice constant. For not too large net baryon density, i.e., when the number of baryons in the system is not much larger than the number of anti-baryons, this leads to numerically accurate results and has been used to demonstrate the existence of a phase transition  between quarks and gluons, the point-like constituents of QCD, and hadrons such as protons and neutrons. Current understanding  is that this transition takes place at temperature T $\approx 157$ MeV and is of 'rapid-cross over' type, i.e., with smooth but rapid change in degrees of freedom between the two phases ~\cite{HotQCD:2014kol,Borsanyi:2010cj}. 

\begin{figure}[htb]
\includegraphics[width=5.5in]{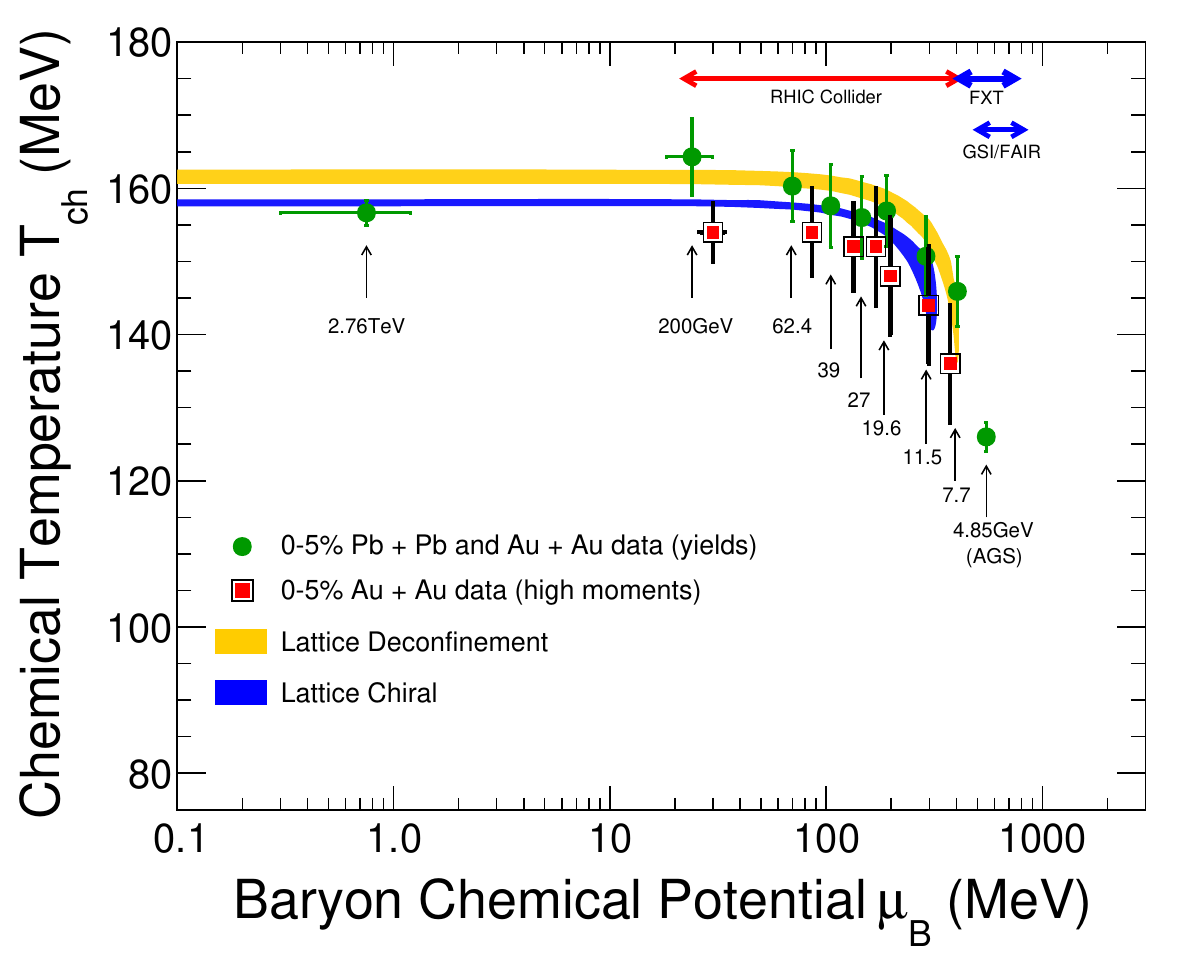}
\caption{Chemical freeze-out temperature $T_{\rm ch}(\mu_{\rm B})$ distribution: Data points are for fixed-target measurements at the AGS and SPS accelerators and for collider measurements at the RHIC and LHC facilities. All data points correspond to 0-5\% central Au-Au or Pb-Pb collisions. Filled circles are extracted from statistical hadronization model (SHM) fits to hadron yields~\cite{STAR:2017sal,Andronic:2017pug} while the red-squares are the results of fits to higher moments~\cite{STAR:2020tga}. The energy coverage of different accelerator complexes is also indicated at the top of the figure. Note that the net-baryon density is closely related to the collision energy: the higher the energy, the smaller the net-baryon density. In phenomenological analyses, the net-baryon density is commonly characterized by the baryon chemical potential $\mu_B$. However, there is no direct experimental measurement of the net-baryon density.}
\label{fig1}
\end{figure}

\begin{figure}[htb]
\includegraphics[width=5.5in]{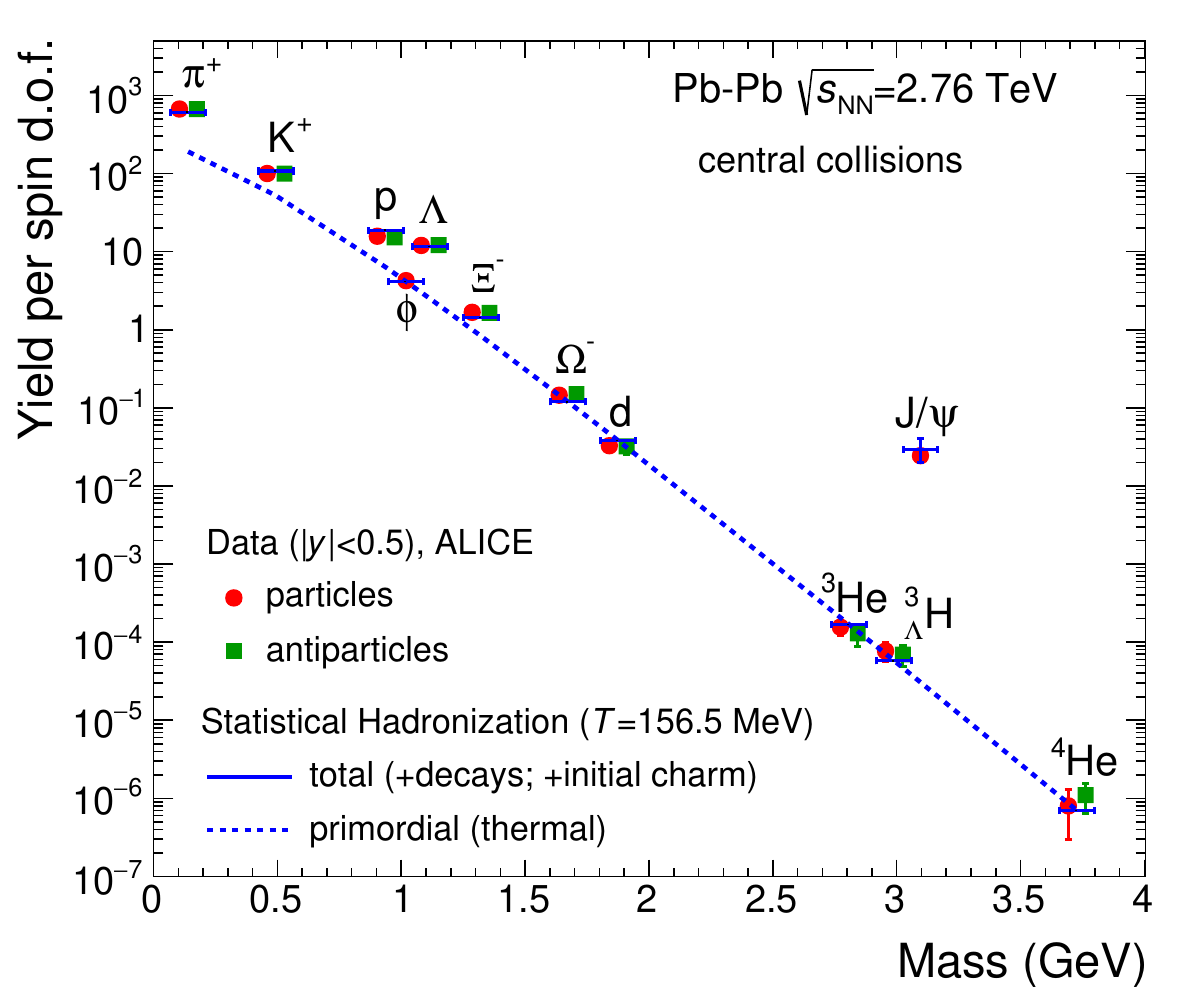}
\caption{Primordial and total (anti-)particle yields, normalized to the
spin degeneracy, as a function of particle mass, calculated with the Statistical Hadronization Model (SHM,SHMc) for
Pb–Pb collisions at $\snn = 2.76$ TeV and compared to data ~\cite{Andronic:2017pug,Andronic:2019wva}. }
\label{fig2}
\end{figure}

\begin{figure}[htb]
\includegraphics[width=5.5in]{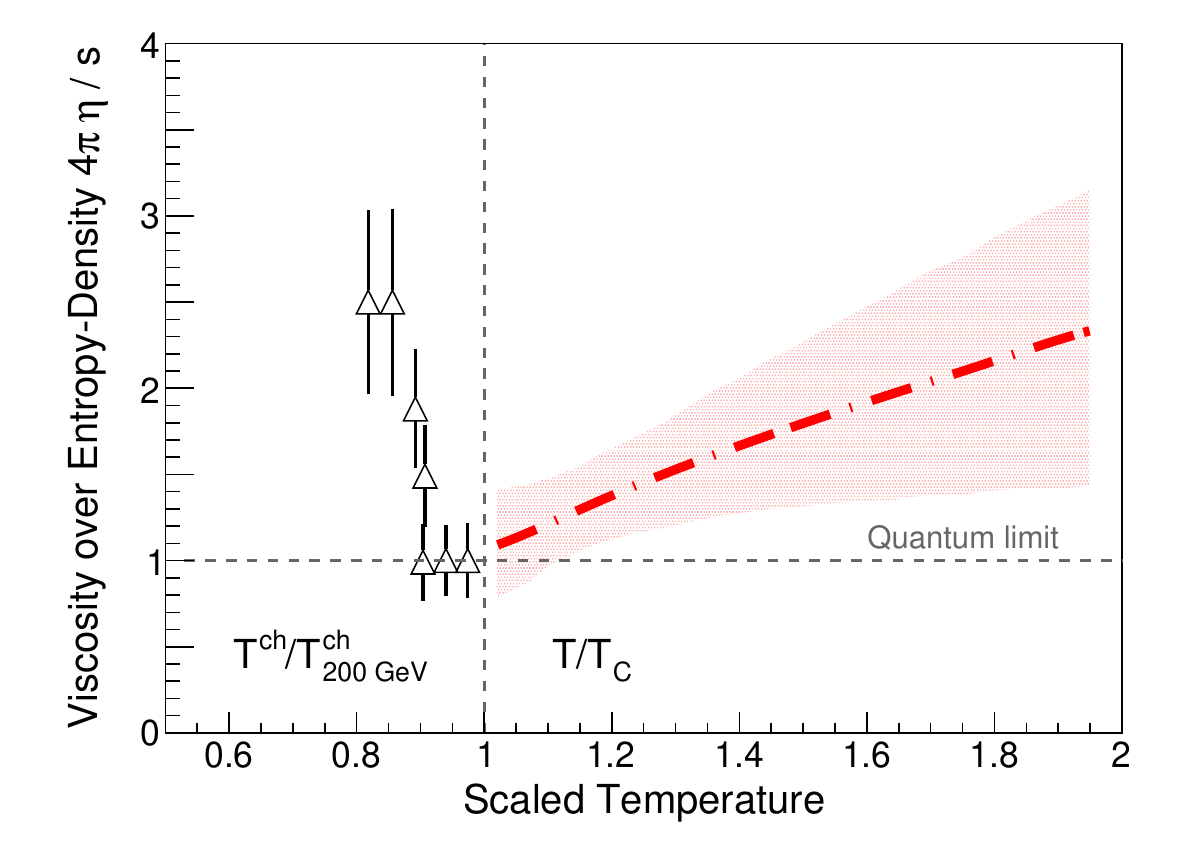}
\caption{Reduced shear viscosity, i.e. shear viscosity-to-entropy-density ratio, $4\pi \eta/s$, shown as a function of the scaled temperature. The horizontal dashed line indicates the quantum lower limit~\cite{Kovtun:2004de}. Left part: the extracted $4\pi \eta/s$ values result from the energy dependence of the measured flow coefficients $v_2$~\cite{STAR:2012och} and $v_3$~\cite{STAR:2016vqt}, displayed here as function of the scaled chemical freeze-out temperature ${\rm T^{ch}/T^{ch}}$(200 GeV). Right part: temperature evolution of $4\pi \eta/s$, extracted from Bayesian analyses~\cite{Bernhard:2019bmu,Xu:2017obm} of data on azimuthal anisotropies measured in relativistic nuclear collisions and analyzed within the framework of relativistic hydrodynamics, for review see~\cite{Braun-Munzinger:2015hba}.}
\label{fig3}
\end{figure}

\begin{figure}[htb]
\includegraphics[width=5.5in]{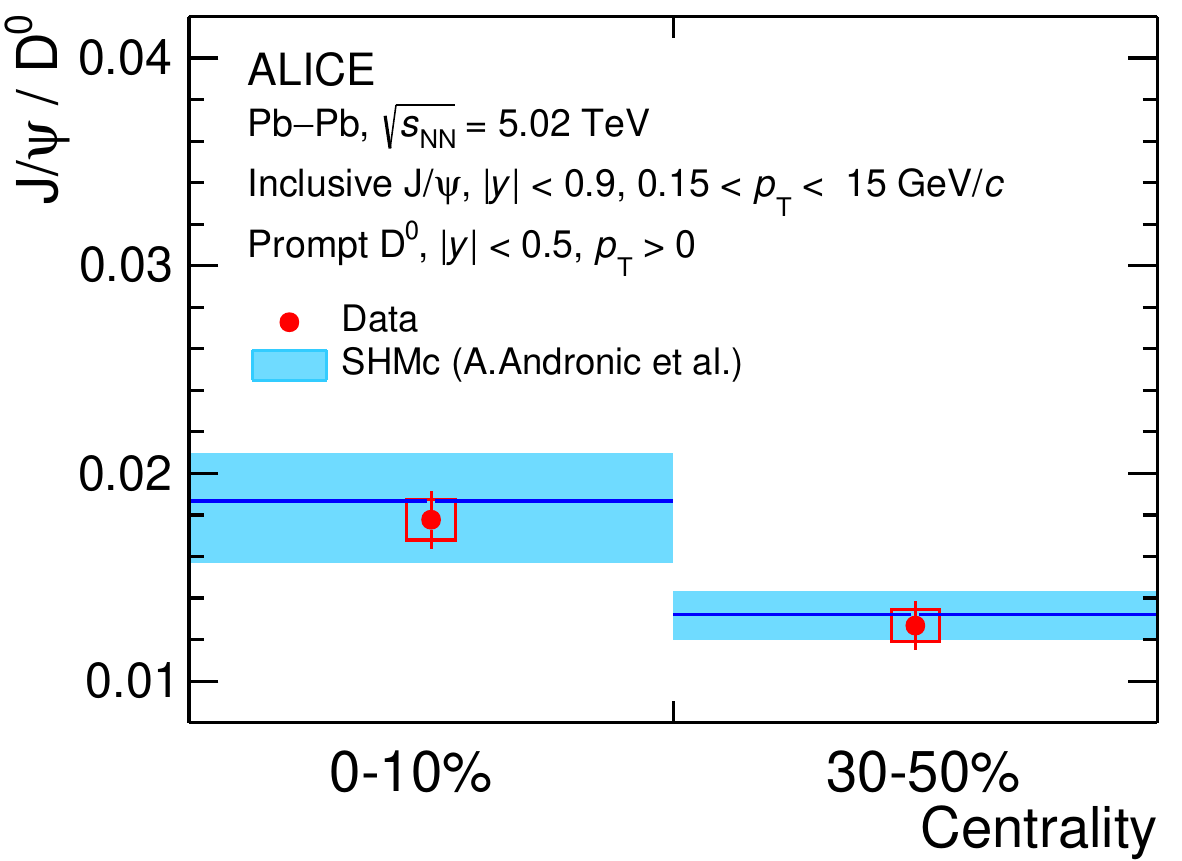}
\caption{   Inclusive J/$\psi$ to D0 yield ratio at $\snn$= 5.02 TeV at mid-rapidity for the 0–10\% and 30–
50\% centrality intervals. Vertical lines and open boxes represent the statistical and systematical uncertainties,
respectively. The measurements are compared with SHMc model predictions  ~\cite{ALICE:2023gco}. For more details see text. }
\label{fig4}
\end{figure}

Experimentally, important information on the QCD phase diagram  was obtained by studying the center-of-mass energy dependence of hadron production in central  Au-Au (AGS, RHIC) and Pb-Pb (SPS and LHC) collisions ~\cite{Andronic:2005yp,Andronic:2017pug}. From this analysis one obtains the density of hadrons as function of the 'chemical freeze-out' temperature T$_{\rm{ch}}$ and baryo-chemical potential $\mu_{\rm{B}}$ that is a measure of the asymmetry between baryons and anti-baryons. The most recent data from the BNL AGS, CERN SPS, RHIC and LHC are displayed in Fig.~\ref{fig1}. 

\begin{marginnote}[]
\entry{Chemical freeze-out}
{Is the stage in a relativistic nuclear collisions when inelastic interactions stop. As consequence, the particle composition of the system is frozen, and the yields of all hadron species remain unchanged thereafter.}
\end{marginnote} 

Two distinct energy regions can be identified. For $\mu_{\rm{B}} \lesssim 100$ MeV, corresponding to $\snn \gtrsim 50$ GeV, the freeze-out points are essentially constant in temperature at T$_{\rm{ch}}\approx 157$ MeV.
We note that lattice QCD calculations predict a small, but finite, curvature for this $\mu_B$ range. This curvature is, however, too small to be measurable with the current experimental precision. The near energy-independence of T$_{\rm{ch}}$ is at first glance surprizing. Note that the energy density as well as the initial temperature of the QGP fireball formed in relativistic nuclear collisions depends significantly on the center-of-mass energy. For a recent compilation see Table 1 in~\cite{Braun-Munzinger:2022bkc, Braun-Munzinger:2025mud}. A natural explanation for the observation of this 'limiting temperature' is that the temperature is stabilized by the QCD phase transition~\cite{Hagedorn:1976ef}.

On the other hand, the chemical potential $\mu_{\rm{B}}$  decreases  strongly with increasing $\snn$, from about 540 MeV at 5 GeV to 0.75 $\pm 0.45$ MeV at 5 TeV ~\cite{Andronic:2017pug,ALICE:2023ulv}. 
The resulting curvature is very well visible in data for $\snn < 50$ GeV~\cite{Andronic:2005yp,Andronic:2017pug}, see Fig.~\ref{fig1}. This is a continuation of the trend predicted by lattice calculations for data at higher energies although lattice calculations are reliable only for $\snn \gtrsim 20$ GeV.

Further analysis of the very precise ALICE data led to an experimental value of temperature parameter which is essentially identical to the theoretical  phase transition temperature of T$_{\rm{pc}} =156.5 \pm 1.5$ MeV~\cite{Andronic:2017pug}, see Fig.~\ref{fig1}. The very close agreement between SHM predictions and data over 9 orders of magnitude, as seen in Fig.~\ref{fig2}, demonstrates that the thermal equilibrium assumption  is  well fulfilled at LHC energy. The special role of the J/$\psi$ meson will be discussed later in the section.

In the following we will summarize what is currently known about the structure of the strongly interacting matter near the phase boundary for relatively small values of $\mu_B < 400$ MeV, where lattice calculations are accurate. For larger $\mu_B$ the unsolved 'sign problem' in lattice QCD prevents model-independent theory predictions. Emphasis will be placed on the crucial viscosity/entropy density ratio and on our present experimental information on deconfinement obtained mostly from the study of hadrons containing charm quarks, see below. 

In Fig.~\ref{fig3} the temperature dependence is shown of the ratio of (shear viscosity)/(entropy density), called reduced viscosity below, as extracted from a hydrodynamic analysis of relativistic nuclear collision data. Although the resulting reduced viscosity values are still sketchy, a clear pattern emerges. Assuming that T$^{ch}$(200 GeV) coincides with the temperature of the QCD phase transition, see the discussion above, the reduced viscosity values first decrease as function of the scaled temperature until the phase transition is reached. For higher temperatures, a smooth increase is observed, see right side of Fig.~\ref{fig3}. This behavior is typical for fluids near a phase transition~\cite{Csernai:2006zz} and provides further evidence for the QCD phase transition. The reduced viscosity values near 1 for temperatures near the phase boundary imply that the 'fireball' produced in relativistic nuclear collisions has the attributes of an ideal  'quantum' fluid~
\cite{Kovtun:2004de}.

The QCD phase transition discovered using lattice QCD methods and experimentally studied with relativistic nuclear collisions is related to the restoration of chiral symmetry in strongly interacting quark-gluon systems at high temperature and density. Inside such systems, one expects also that the color confinement is lifted. For ordinary hadrons confinement implies  that the quark and gluon constituents  are not observable as free particles. In contradistinction, color is 'deconfined' in a QGP droplet, implying that quarks and gluons can travel freely within the volume of the droplet. 

This phenomenon has recently been observed directly~\cite{Gross:2022hyw,ALICE:2023gco}. The idea for this measurement came out of the realization that, in relativistic nuclear collisions, the charm quark content of the formed QGP droplet is, to good accuracy, conserved during the fireball evolution~\cite{Andronic:2006ky}.  Charm hadron yields should be strongly enhanced if the charm quarks produced copiously in initial hard collisions are effectively thermalized in the hot medium~\cite{Braun-Munzinger:2000csl,Andronic:2017pug}. Strong evidence for such charm quark thermalization has recently been observed
~\cite{Andronic:2019wva,Andronic:2021erx,ALICE:2021rxa}. The corresponding enhancement factor g$_c$  is not a free parameter (in thermal parlance a fugacity) but determined  directly from the measured total open charm cross section. This results for central Pb-Pb collisions at $\snn = 5$ TeV in $\rm{g}_c =30$. This is directly   visible in Fig.~\ref{fig2}: because the J/$\psi$ contains two charm quarks its enhancement factor over the purely thermal yield should be  $\rm{g}_c^2 = 900$, exactly as observed. The resulting 'charm version' of the SHM is called SHMc below and is a modified but still thermal version of SHM.

From the thermalized charm quarks,  charm mesons can only be 'assembled' effectively at the QCD phase transition if all charmed quarks are deconfined within the size of the QGP. The ratio of the yields of hadrons with one charm quark (D$_0$) and two charm quarks (J/$\psi$) then is a measure of the charm quark density n$_c$. 
The ALICE collaboration has recently measured the yield of D$_0$ and J/$\psi$ mesons in Pb-Pb collisions along with $\rm{V}_{\rm{SHM}}$ at LHC energy. 

The result is displayed in Fig.~\ref{fig4} where one sees that  the measured ratio under the assumption of deconfinement is in very good agreement with  SHM$_c$ prediction assuming complete deconfinement.

Clearly significant progress has been made in our understanding of the QCD phase structure for small baryon chemical potential $\mu_B \lesssim 400$ MeV. For larger values of $\mu_B$ \emph{ab initio} QCD calculations are not available and one has to resort to predictions using various effective models. 
In this context, particular efforts were recently undertaken to establish 'hadronic baseline' models without a critical endpoint, see below.

Experimentally, particle production measurements have been, over the past decade, extended to include not just mean values (1st-order moments) but also provide predictions for higher-order moments. For good recent summaries see~\cite{Bzdak:2019pkr, Luo:2017faz, Rustamov:2022hdi}. Such fluctuation measurements are particularly motivated by the search for a critical endpoint in the QCD phase diagram~\cite{Stephanov:1999zu,Stephanov:2004wx}. Searches for signatures of a critical endpoint have been undertaken within the framework of lattice QCD. No signs of a critical point have been observed to-date (Nov. 2025) for baryo-chemical potentials $\mu_B < 450$ MeV~\cite{Borsanyi:2025dyp}. An extensive experimental campaign has been just completed with the beam energy scan (BES) program at the BNL RHIC facility. 

In this review we will focus on event-by-event proton number fluctuation measurements in the energy range from $4.8 < \snn  < 5500 $ GeV with special emphasis on the low energy BES range, $\snn  \leq 200 $ GeV. One of the aims is to provide a basis for further studies of the QCD phase transition in an energy range where a possible critical endpoint~\cite{Stephanov:1999zu,Sorensen:2024mry} might lie. 

We note that the QCD phase transition is, in its nature,  not  similar to the 'liquid-gas' phase transition but contains completely new elements which are outside the realm of classical nuclear physics. An important new aspect is the 'hadronization' process which involves a change of degrees of freedom by roughly a factor of 3.5 between partons on one side and hadrons on the other side of the transition. At (nearly) constant entropy this leads to a significant volume increase (by a similar factor) resulting in a corresponding time delay during hadronization. At the same time,  chiral symmetry is broken as one crosses the transition region from the high temperature side. Whether and how the presence of this  phase transition or of a possible critical point in it can influence the interactions in the dominantly baryonic matter at the low-temperature side of the transition is an important but completely open question. We will come back to this during the discussion of experimental fluctuation results below.

 Importantly, in the vicinity of the QCD critical point, the detailed microscopic dynamics of QCD, including hadronization processes and the quark–gluon substructure, should become irrelevant for the critical behavior. Instead, the long-distance physics is governed by a single scalar collective mode, analogous to density fluctuations near the liquid–gas critical point. As a consequence, the QCD critical point is expected to belong to the three-dimensional Ising $Z(2)$ universality class, leading to universal scaling laws that are largely insensitive to the microscopic realization of the medium. As a result, baryon-number fluctuations couple directly to the critical mode and inherit its universal scaling behavior. This is  why fluctuations of conserved quantities, in particular proton-number fluctuations as a proxy for baryon-number fluctuations, are key observables in the experimental search for the QCD critical point.

Particularly important for such studies is to establish a theoretical baseline for models without any criticality~\cite{Braun-Munzinger:2020jbk,Vovchenko:2022nhb, Friman:2025swg}. An important contribution also comes from hadronic multiple collision models such as UrQMD~\cite{Sombun:2017bxi} and SMASH~\cite{Elfner:2025ojd}. The latest results from the BES experiment will be confronted with predictions from various models, and systematically analyzed for possible indications of critical phenomena.

\begin{marginnote}[]
\entry{Baryon chemical potential, $\mu_{B}$ }{Is the cost in Helmholtz free energy ($F$) of changing the baryon number ($N_{B}$) of a thermal system by one, at fixed temperature and volume, $\mu_{B}=\left(\frac{\partial F}{\partial N_{B}}\right)_{T,V}$.
It appears explicitly in the Grand Canonical Ensemble (GCE) distribution functions for baryons and antibaryons.}
\entry{Net-baryon density}{Defined as the difference between baryon and antibaryon number densities: $n_{B}^{net} = n_{B} - n_{\bar{B}}$. As $\mu_{B}$ increases $n_{B}$ grows, while $n_{\bar{B}}$ decreases, so $n_{B}^{net}$ increases. At $\mu_{B}$ = 0, $n_{B}^{net}$ = 0. In the Boltzmann limit, $n_{B}^{net}\propto sinh(\frac{\mu_{B}}{T})$
}
\end{marginnote}

\section{Deciphering the QCD phase diagram with multiplicity distributions}
Multiplicity distributions of hadrons, measured experimentally, offer a unique opportunity to quantify the properties of QCD matter created in head-on (central) collisions of heavy nuclei. The Statistical Hadronization Model (SHM) provides a framework to calculate cumulants of these distributions. The essential idea of SHM is simple but powerful: once the fireball produced in heavy-ion collisions reaches the QCD transition, hadrons are formed according to the statistical laws of equilibrium thermodynamics. This process is referred to as chemical freeze-out.  

At chemical freeze-out, hadron yields are determined solely by a few global parameters: the temperature $T$, the baryon chemical potential $\mu_{B}$, and the fireball volume $V$. The central task of SHM is to determine the set of these parameters that simultaneously describes the measured yields, i.e.,  the first cumulants of the multiplicity distributions of all measured hadron species. The starting point of this procedure is the construction of the partition function, from which all thermodynamic quantities, including  cumulants, can be calculated. In SHM the partition function of the system is approximated by that of an ideal gas composed of all stable hadrons and resonances, hence also referred to as the Hadron Resonance Gas (HRG). Interactions can be taken into account via the S-matrix correction~\cite{Andronic:2018qqt,Cleymans:2020fsc} if experimental information on relevant phase shifts is available.  The grand-canonical partition function of the HRG is constructed as

\begin{equation}
\ln \mathcal{Z}(T,V,\mu_{B}) = 
\sum_{i} \pm \frac{g_{i}V}{(2\pi)^{3}} 
\int d^{3}p \, \ln \!\left(1 \pm e^{-(E_{i}-\mu_{i})/T}\right),
\end{equation}
where the upper (lower) sign corresponds to fermions (bosons), the sum runs over all hadron and resonance species $i$, with spin degeneracy factor $g_{i}$ and total energy 
\begin{equation}
E_{i} = \sqrt{p^{2}+m_{i}^{2}},
\end{equation}
The chemical potential of each hadron is given by
\begin{equation}
\mu_{i} = B_{i}\mu_{B} + Q_{i}\mu_{Q} + S_{i}\mu_{S} + C_{i}\mu_{C},
\end{equation}
where, $B_{i}$, $Q_{i}$, $S_{i}$, and $C_{i}$ denote the baryon number, electric charge, strangeness, and charm quantum numbers of species $i$, respectively. Although in principle each conserved charge introduces a corresponding chemical potential, in practical SHM applications only $\mu_{B}$ is treated as independent, while $\mu_{Q}$, $\mu_{S}$, and $\mu_{C}$ are fixed by physical constraints. 

All thermodynamic observables can be expressed as derivatives of $\ln \mathcal{Z}$. For instance, the pressure reads

\begin{equation}
    p \equiv T\frac{\partial \ln \mathcal{Z}(T,V,\mu_{B})}{\partial V} = \frac{T}{V}\ln \mathcal{Z}(T,V,\mu_{B}) = \sum_{i}p_{i},
\end{equation}
where $p_{i}$ is the partial pressure of the ideal Fermi or Bose gas of species $i$. Similarly, mean values (first-order cumulants), variances (second-order cumulants), and higher-order fluctuations of particle numbers follow directly from the HRG partition function. These quantities can be directly compared to the corresponding cumulants of the experimentally measured multiplicity distributions, thereby providing a powerful tool to explore the properties of QCD matter and to chart its phase structure.

\subsection{Cumulants and factorial cumulants of multiplicity distributions}
\label{sec_cumulants}

Cumulants and factorial cumulants provide a systematic and physically transparent framework for quantifying the structure of multiplicity distributions and identifying dynamical effects, such as the nature of interactions among particles under study, underlying particle production. For a random variable $N$ with probability distribution $P(N)$, the cumulants $\kappa_{n}$ are defined through the logarithm of the moment-generating function
\begin{equation}
    K(t) = \ln \langle e^{tN}\rangle, \qquad
    \kappa_{n}=\left.\frac{d^{n}K(t)}{dt^n}\right|_{t=0},
    \label{eq_cumulants}
\end{equation}
where $t$ does not correspond to any physical observable; instead, it serves as a mathematical expansion parameter that allows all cumulants to be obtained through derivatives of $K(t)$ at $t=0$. 
Ratios of cumulants such as $\kappa_{2}/\kappa_{1}$, $\kappa_{3}/\kappa_{2}$, and $\kappa_{4}/\kappa_{2}$ are widely used in experimental analyses because they reduce sensitivity to overall normalization and provide comparisons to baseline expectations. In particular, near a QCD critical point, higher-order cumulants are predicted to exhibit large, non-monotonic behavior and even sign changes as function of energy due to the growth of the correlation length. Thus, precise measurement of $\kappa_{3}$ and beyond is considered one of the most sensitive experimental probes of critical phenomena in strongly interacting matter.

While cumulants provide a compact summary of statistical properties, they mix contributions from correlations of different orders. To disentangle these effects, factorial cumulants, also known as integrated correlation functions, offer a more fundamental representation of particle production mechanisms. Analogously to Eq.~\ref{eq_cumulants}, they are defined as derivatives of the factorial cumulant generating function
\begin{equation}
    G(t) = \ln \langle (1+t)^N \rangle, \qquad
    C_{n}=\left.\frac{d^{n}G(t)}{dt^n}\right|_{t=0}.
    \label{eq_factcumulants}
\end{equation}

The factorial cumulant $C_{n}$ quantifies the strength of genuine $n$-particle correlations. The first two factorial cumulants read
\begin{equation}
    C_{1}=\langle N\rangle, \qquad
    C_{2}=\langle N(N-1)\rangle - \langle N\rangle^2.
\end{equation}

\begin{marginnote}[]
\entry{Factorial cumulants}
{factorial cumulants $C_{n}$, also known as integrated correlation functions, isolate genuine $n$-particle correlations and therefore constitute particularly suitable observables for probing critical phenomena.}

\end{marginnote}

The second factorial cumulant $C_{2}$ isolates the genuine two-particle correlation, because the term $\langle N(N-1)\rangle$ contains both statistical and dynamical pair correlations, while the subtraction of $\langle N\rangle^2$ removes the part arising from uncorrelated (Poissonian) particle emission. As a result, $C_{2} = 0$ for a purely independent production process.

In a similar way, $C_{3}$ isolates the genuine three-particle correlation by systematically removing all contributions arising from independent emission and from lower-order (pairwise) correlations, ensuring that $C_{3} = 0$ for any process governed solely by independent particle production. The same hierarchical mechanism applies to higher orders: each $C_{n}$ subtracts all combinations of correlations of order $<n$, leaving only the genuine $n$-particle connected correlation.

Thus, for a Poisson distribution, all factorial cumulants of order $n \ge 2$ vanish identically, reflecting the complete absence of genuine multi-particle correlations. Consequently, any non-zero value of $C_{n}$ directly indicates true $n$-particle correlations arising from, for example, resonance decays, clustering phenomena, conservation laws, or critical fluctuations.

The relation between factorial cumulants $C_{n}$ and ordinary cumulants $\kappa_{n}$ 
is mediated by the signed Stirling numbers of the first kind $s(n,j)$

\begin{equation}
    C_{n} = \sum_{j=1}^{n} s(n,j)\,\kappa_{j},
    \label{eq_stirling2}
\end{equation}
For the first four cumulants Eq.~\ref{eq_stirling2} yields
\begin{eqnarray}
    C_{1} &=& \kappa_{1} = \langle N\rangle,\nonumber\\
    C_{2} &=& -\kappa_{1} + \kappa_{2},\\
    C_{3} &=& 2\kappa_{1} - 3\kappa_{2} + \kappa_{3},\nonumber\\
    C_{4} &=& -6\kappa_{1} + 11\kappa_{2} - 6\kappa_{3} + \kappa_{4}\nonumber.
\end{eqnarray}

The inverse relation is given by the Stirling numbers of the second kind, $S(n,j)$
\begin{equation}
    \kappa_{n} = \sum_{j=1}^{n} S(n,j)\,C_{j},
\end{equation}
leading to
\begin{eqnarray}
    \kappa_{1} &=& C_{1} = \langle N\rangle, \nonumber\\
    \kappa_{2} &=& C_{1} + C_{2},\\
    \kappa_{3} &=& C_{1} + 3C_{2} + C_{3},\nonumber\\
    \kappa_{4} &=& C_{1} + 7C_{2} + 6C_{3} + C_{4} \nonumber.
\end{eqnarray}

The combined use of ordinary and factorial cumulants provides a powerful and complementary toolkit for disentangling dynamical correlations, conservation-law effects, resonance decays, and possible critical phenomena in relativistic nuclear collisions.

\subsection{First-order cumulants}
The primordial mean multiplicity (first-order cumulant) of hadron species $i$ is obtained as
\begin{equation}
\langle N_i^{prim}\rangle = \frac{\partial\ln\mathcal{Z}(T,V,\mu_{B})}{\partial (\mu_i/T)}.
\label{eq-mean}
\end{equation}
The total mean multiplicity measured experimentally also includes contributions from resonance decays 

\begin{equation}
\langle N_i^{tot} \rangle = \langle N_i^{prim}\rangle 
+ \sum_{j} \langle n_{i|j} \rangle \, \langle N_j^{prim} \rangle,
\end{equation}
where $\langle n_{i|j} \rangle$ denotes the average number of hadrons $i$ produced in the decay of resonance $j$. In practical SHM/HRG calculations, these decay contributions are systematically included by using the experimentally known branching ratios as tabulated in the Particle Data Group (PDG)~\cite{ParticleDataGroup:2024cfk}. The quantity $\langle n_{i|j} \rangle$ also accounts for contributions resulting from sequential decay chains via intermediate lower-mass resonances. By fitting the calculated total yields $\langle N_i^{tot} \rangle$ to experimental data, the global fireball parameters at freeze-out can be determined.  In Fig.~\ref{fig2}, the primordial yields $\langle N_i^{prim} \rangle$ and the total yields $\langle N_i^{tot} \rangle$ obtained in the SHM/HRG framework are compared with the experimentally measured hadron multiplicities at mid-rapidity by the CERN ALICE experiment. The comparison yields the freeze-out parameters $T_{ch}^{\mathrm{ALICE}} = 156.5 \pm 1.5$ MeV, $\mu_B = 0.7 \pm 3.8$ MeV, and $V = 5180 \pm 410$ fm$^3$. Recently, new and precise measurements by the ALICE collaboration~\cite{ALICE:2023ulv} have led to a significant improvement in accuracy for the baryon chemical potential:   $\mu_B = 0.71 \pm 0.45$ MeV. This implies that the fireball formed at LHC energy is essentially net-baryon free, very similar to the matter in the early universe at $10 \mu s$ after the big bang. It should be noted that, unlike the temperature and baryon chemical potential, the volume parameter is calculated from the measured yield of charged particles  rather than treated as a free fit parameter.

When the number of hadrons at freeze-out is small, they are treated in the canonical ensemble, enforcing event-by-event exact conservation of the relevant quantum number. For a hadron $i$ carrying a conserved quantum number $q_{i}$, the canonical suppression factor $f_{q_{i}}^{can}$ quantifies how much the exact conservation of that quantum number reduces its mean multiplicity compared to the grand-canonical (GC) expectation. It should be noted that the canonical suppression factor does not affect particles carrying hidden quantum numbers, such as the $\phi$ meson in the case of strangeness or the $J/\psi$ meson in the case of charm. These states consist of a quark-antiquark pair of the same flavor. Therefore, their production does not require the creation of a corresponding antiparticle to ensure global conservation, and their yields remain unaffected by canonical suppression. 

Charm quarks are produced mainly in
initial hard scatterings, and their total abundance stays essentially constant during the evolution before hadronization, see also the discussion of Fig.~\ref{fig4} above. In SHM this is accounted for by introducing a charm fugacity parameter $g_{c}$, which is determined from measurements of the total open-charm cross section.
The introduction of $g_{c}$, however, does not imply the presence of off-equilibrium effects since the charm quarks  thermalize kinetically, as discussed in the introduction. Rather, $g_{c}$  enforces charm conservation, reflecting the fixed number of pairs produced in the initial hard scatterings. Once the system hadronizes, the  $c$ and $\bar{c}$ quarks are distributed statistically into different charmed hadrons following the standard SHM procedure.

\begin{marginnote}[]
\entry{Direct evidence for deconfinement}{The empirical success of the SHMc in reproducing the observed $J/\psi$ to $D^0$ ratio constitutes direct experimental evidence for deconfinement of charm quarks.}
\end{marginnote}

Contrary to the canonical suppression factor $f_{C_{i}}^{can}$, the fugacity factor $g_{c}$ affects both particles with hidden and open charm quantum numbers.  Therefore, the empirical success of the SHMc in reproducing the observed $J/\psi$ to $D^0$ ratio implies that charm quarks were deconfined prior to hadronization. Indeed, this agreement indicates that charm quarks, although not in chemical equilibrium, become mobile in a deconfined medium and subsequently hadronize statistically at the QCD phase boundary. In contrast, light quarks are produced copiously in the collision and  reach  chemical and kinetic equilibrium during the partonic phase of the hot fireball. 

\subsection{Higher-order cumulants}
The higher-order cumulants of particle number distributions are obtained analogously to Eq.~\ref{eq-mean}. 

Sometimes, however, only the full QCD partition function is available, as in lattice QCD calculations discussed in the following sections. In such cases, derivatives of the partition function can be taken directly with respect to the chemical potential $\mu_{q}$, which controls the conservation of the corresponding conserved charge $q$ on average

\begin{equation}
    \kappa_{n}(N_{q}) = \left. \frac{\partial^n\ln\mathcal{Z}(T,V,\mu_{B})}{\partial (\mu_q/T)^n}\right|_{T,\, V = \mathrm{const}, \mu_{q} = 0} = VT^{3}\chi_{n}^{q},
    \label{eq_chi}
\end{equation}
where $\chi_{n}^{q}$ denote the generalized susceptibilities, defined as derivatives of the dimensionless pressure ($p/T^4$) with respect to the dimensionless chemical potentials ($\mu_q/T$). 

Eq.~\ref{eq_chi} lies at the heart of fluctuation analyses, as it establishes a direct connection between experimentally measurable observables, such as the cumulants of conserved-charge number distributions and the generalized susceptibilities, which encode essential information about the QCD equation of state through its relation to the thermodynamic pressure. 

\subsection{Lattice QCD predictions}
\label{secLQCD}
Lattice Quantum Chromodynamics (lattice-QCD) provides first-principles framework for determining the thermodynamic properties of strongly interacting matter at finite temperature and vanishing or moderate baryon chemical potential. In the context of relativistic heavy-ion collisions, lattice-QCD calculations establish equilibrium benchmarks for bulk thermodynamic observables, as well as for fluctuations and correlations of conserved charges, under well-controlled theoretical conditions.

Direct lattice simulations at finite baryon chemical potential are obstructed by the fermion sign problem. Consequently, lattice-QCD predictions at nonzero $\mu_B$ are obtained through systematic Taylor expansions of thermodynamic observables around $\mu_B = 0$. State-of-the-art calculations currently include expansion coefficients up to sixth or eighth order, enabling controlled extrapolations to the range of baryon chemical potentials relevant for the Beam Energy Scan program.

The primary lattice-QCD observables relevant for fluctuation measurements are the generalized susceptibilities of conserved charges,
\begin{equation}
\chi^{BQS}_{ijk}(T,\mu_B,\mu_Q,\mu_S)
=
\frac{\partial^{\,i+j+k}}{\partial(\mu_B/T)^i\,\partial(\mu_Q/T)^j\,\partial(\mu_S/T)^k}
\left(\frac{p(T,\mu_B,\mu_Q,\mu_S)}{T^4}\right),
\end{equation}
where $p$ denotes the pressure, and $B$, $Q$, and $S$ correspond to baryon number, electric charge, and strangeness, respectively. In the grand-canonical ensemble, these susceptibilities are directly related to the cumulants of event-by-event distributions of conserved charges (cf.~Eq.~\ref{eq_chi}).

The different sensitivities of baryon number cumulants of various orders to the QCD crossover are closely tied to the smallness of the physical light-quark masses. In the chiral limit, QCD exhibits a second-order phase transition governed by O(4) universality~\cite{Karsch:2003jg}. For small but finite quark masses, this transition is replaced by a smooth crossover~\cite{Aoki:2006we}; nevertheless, remnants of the underlying chiral critical dynamics persist and leave characteristic imprints on higher-order fluctuations~\cite{Ejiri:2005wq, Friman:2011pf}.

In this situation, lower-order susceptibilities are dominated by regular, non-critical contributions to the thermodynamic potential. Second-order baryon number fluctuations therefore vary smoothly with temperature and remain well described by HRG prediction with the ideal gas equation of state~\cite{ALICE:2019nbs}. Fourth-order cumulants already show deviations from Hadron Resonance Gas expectations in the transition region~\cite{HotQCD:2012fhj, Borsanyi:2011sw}. However, within O(4) universality the singular contribution to fourth-order fluctuations is subleading and does not generate a unique or universal structure. Consequently, deviations observed at fourth order cannot be interpreted as direct evidence for the chiral crossover~\cite{Bazavov:2012vg}.

In contrast, sixth-order baryon number cumulants are the lowest-order observables that receive a leading contribution from the singular part of the free energy. Lattice-QCD calculations show that these cumulants exhibit qualitatively distinct behavior in the crossover region, including a characteristic sign change and strong non-monotonic temperature dependence, which persist for physical quark masses. This establishes sixth-order fluctuations as the first thermodynamic observables that provide unambiguous sensitivity to the chiral dynamics underlying the QCD crossover.

Lattice QCD provides important constraints on the existence and possible location of a critical point in the QCD phase diagram. Quantitatively, present lattice-QCD results exclude the presence of a critical point for baryon chemical potentials up to $\mu_{B}/T \sim 3$~\cite{Bazavov:2020bjn, Borsanyi:2021sxv}. If a critical point exists, it must therefore be located at larger baryon chemical potentials, beyond the region that can currently be accessed by controlled lattice extrapolations.

A direct quantitative comparison between lattice-QCD results and experimental measurements is, however, not straightforward. Lattice calculations are performed in the grand-canonical ensemble, where conserved charges fluctuate freely and global conservation laws are not explicitly enforced. In contrast, in heavy-ion collisions baryon number, electric charge, and strangeness are conserved exactly on an event-by-event basis. Furthermore, lattice-QCD calculations are carried out at fixed volume, whereas the effective system volume in experiments fluctuates from event to event, generating additional contributions to the measured cumulants.

Experimental measurements are also subject to finite detector acceptance, which modifies the impact of global conservation laws on the observed fluctuations. In particular, increasing acceptance leads to a stronger suppression of fluctuations induced by charge conservation, especially for baryon number. These acceptance effects must therefore be carefully accounted for when confronting lattice-QCD predictions with experimental data.

An additional limitation arises from the fact that lattice-QCD calculations are formulated for conserved charge fluctuations, such as net-baryon number, while experimental measurements typically access net-proton fluctuations as a proxy. Lattice-QCD predictions for fluctuations of individual baryon species are not available. By contrast, experiments can measure proton and antiproton multiplicity distributions separately, as well as their mixed cumulants. As a result, lattice-QCD calculations alone are insufficient for a quantitative interpretation of experimental fluctuation measurements. Complementary baseline calculations that explicitly incorporate conservation laws, volume fluctuations, acceptance effects, and the relation between net-baryon and net-proton observables are therefore essential.

\section{Baselines for fluctuations measurements}

For a quantitative interpretation of experimental measurements, it is essential to establish well-defined reference baselines. A natural baseline is provided by the Hadron Resonance Gas (HRG) model formulated within the Grand Canonical Ensemble (GCE) in the Boltzmann approximation. In this limit, where quantum statistical effects are negligible, all cumulants of the multiplicity distribution for a single hadron species reduce to its mean value, and the multiplicity distributions are governed by Poisson statistics.
For net-particle number distributions, the situation corresponds to the difference between two statistically independent Poisson processes, leading to a Skellam distribution. The cumulants of particle and net-particle distributions are then given by 

\begin{equation}
     \kappa_{n}(N_{q})=\langle N_{q}\rangle,
      \label{eq_baseline_skellam_1}
\end{equation}
\begin{equation}
    \kappa_{n}(N_{q}-N_{\bar{q}})=\langle N_{q}\rangle + (-1)^n\langle N_{\bar{q}}\rangle
     \label{eq_baseline_skellam}
\end{equation}
   
\subsection{Baselines with exact conservation laws}
Although the baseline expressions given in 
Eqs.~\ref{eq_baseline_skellam_1} and~\ref{eq_baseline_skellam} are formally simple, they cannot be directly compared to experimental measurements. This is because these baselines are derived under the assumption that baryon number conservation is satisfied only on average over the full phase space, as implemented in the grand canonical ensemble. In reality, strong interactions conserve $B$, $Q$, and $S$ exactly in every single event. The correct statistical framework for describing fluctuations under exact conservation is, therefore, the canonical ensemble (CE), in which the conserved charges of the system are fixed event-by-event. The consequences of exact conservation are dramatic:

\begin{itemize}
    \item In the full ($4\pi$) acceptance, any conserved charge, such as baryon number, electric charge, or strangeness, is completely determined by the quantum numbers of the colliding nuclei and takes the same value in every event. As a result, event-by-event fluctuations of these globally conserved quantities vanish identically. This behavior is the defining characteristic of the canonical ensemble.
    \item In a finite, but not too small acceptance, the total fireball still belongs to the canonical ensemble with exactly fixed global conserved charges. The observed subsystem, however, is open: it can exchange particles carrying these charges with the unobserved part of the system. As a result, fluctuations of quantities such as the net-baryon number within the acceptance window arise solely from particles migrating across the acceptance boundary and are strongly suppressed compared to the grand-canonical (Skellam) expectation. These fluctuations are fully accounted for within the canonical framework. Exact global conservation combined with finite acceptance can be calculated analytically as well as in Monte Carlo simulations. 
    
    For example, the second-order cumulant of net-baryon number fluctuations, $\kappa_{2}(B-\bar{B})$, is suppressed relative to the corresponding Skellam expectation $\langle n_{B} + n_{\bar{B}}\rangle$~\cite{Braun-Munzinger:2020jbk}:

    \begin{equation}
\frac{\kappa_{2}(B-\bar{B})}{\langle n_{B}+n_{\bar{B}}\rangle}= 1- \beta,
\label{eq_can_kappa2}
\end{equation}
where the suppression factor $\beta > 0$ is given by

\begin{equation}
\beta = \frac{\alpha_{B}\langle n_{B}\rangle + \alpha_{\bar{B}}\langle n_{\bar{B}}\rangle}{\langle n_{B}+n_{\bar{B}}\rangle} + \frac{(\alpha_{B}-\alpha_{\bar{B}})^{2}}{\langle n_{B}+n_{\bar{B}}\rangle}\left(z^2-\langle N_{B}\rangle \langle N_{\bar{B}}\rangle\right),
\label{eq_beta}
\end{equation}
where $\alpha_{B}$ and $\alpha_{\bar{B}}$ denote the accepted fractions of baryons and antibaryons, respectively, and $z$  is the single-baryon partition function. Similar expressions can be derived for arbitrarily higher-order cumulants as well.

At low collision energies, where the presence of antibaryons can be neglected, one may take
$\langle N_{\bar{B}} \rangle = \langle n_{\bar{B}} \rangle = z = \alpha_{\bar{B}} = 0$.
In this limit the canonical suppression factor $\beta$ depends on a single parameter, $\alpha_{B}$, i.e. $\beta = \alpha_{B}$~\cite{Braun-Munzinger:2020jbk}. Remarkably, the same single–parameter form also applies at LHC energies: owing to baryon–antibaryon symmetry at mid-rapidity, one may impose
$\alpha_{B} = \alpha_{\bar{B}}$ when evaluating Eq.~\ref{eq_beta}~\cite{ALICE:2019nbs}.

For all other intermediate collision energies, the full expressions containing both $\alpha_{B}$ and $\alpha_{\bar{B}}$ must be used.

    \item In the limit of very small acceptance, fluctuations become Poissonian~\cite{LandauStatPhys}, but this behavior is not a consequence of the Grand Canonical ideal-gas picture; rather, it results from the trivial Bernoulli sampling of only a few particles from a finite system. Hence, one must clearly distinguish this case from the Grand Canonical Ensemble, where Poissonian fluctuations arise due to genuine thermodynamic reasons and the absence of correlations in an ideal gas.
\end{itemize}

\subsection{Baselines including interactions}
The next step toward establishing more realistic baselines involves a phenomenological modification of the equation of state (EoS) underlying the HRG model. While retaining the canonical ensemble framework to ensure exact baryon number conservation, baryonic interactions can be incorporated in order to go beyond the ideal-gas approximation and introduce a non-ideal EoS.

A concrete implementation of this idea was presented in Ref.~\cite{Vovchenko:2021kxx}, where viscous hydrodynamic simulations were supplemented with a correction factor applied to the single-particle distribution function entering the Cooper--Frye particlization prescription. This factor accounts for deviations of the interacting HRG from the ideal-gas limit and depends on the particular interaction scheme employed within the HRG. 

In Ref.~\cite{Vovchenko:2021kxx}, repulsive baryonic interactions were modeled phenomenologically by assigning finite excluded volumes to baryons. In this approach, baryons are effectively endowed with hard-core radii, which prevents them from occupying the same spatial volume. This prescription introduces a geometric constraint in configuration space, such that the available volume for placing additional baryons becomes reduced as the baryon density increases. As a consequence, non-trivial correlations among baryons arise from purely geometric packing effects. These correlations reduce the accessible phase-space volume and, in turn, modify thermodynamic quantities such as pressure, particle-number susceptibilities, and higher-order cumulants of conserved charges. In the following discussion, we refer to this framework as hydrodynamics with excluded volume (Hydro-EV).

A different approach was proposed in Ref.~\cite{Friman:2025swg}, where both repulsive and attractive interactions among baryons are introduced, motivated by equilibrium statistical mechanics. Specifically, a joint probability density function for baryon pairs in rapidity space is defined according to the Boltzmann--Gibbs form
\begin{equation}
P(y_1, y_2) \propto \exp[-E(y_{1},y_{2})/T].
\end{equation}
Within this construction, generated baryon configurations corresponding to low-energy states are statistically favored, whereas those associated with high-energy configurations are suppressed.

To characterize the nature of correlations, the interaction energy $E(y_1, y_2)$ is decomposed into two contributions:
\begin{itemize}
    \item \textbf{A short-range repulsive term}, which rapidly decreases with increasing separation and prevents baryons from approaching each other too closely~\cite{Friman:2025swg}.
    \item \textbf{A long-range attractive term}, which increases with separation and therefore acts as a confining contribution~\cite{Friman:2025swg}.
\end{itemize}

The parameters of the model, governing the strengths of the attractive and repulsive components, are determined via the Metropolis sampling procedure introduced in Ref.~\cite{Braun-Munzinger:2023gsd}.

Clearly, the repulsive interactions introduced via the excluded-volume prescription are implemented in coordinate space, whereas both attractive and repulsive interactions in Ref.~\cite{Friman:2025swg} are implemented in momentum space. Remarkably, despite these fundamentally different realizations of baryonic interactions, both approaches successfully and simultaneously reproduce the experimentally measured proton-number cumulants reported by the STAR Collaboration. This consistency holds down to collision energies of $\sqrt{s_{NN}} \approx 7.7~\mathrm{GeV}$. At sufficiently high collision energies, such correlations naturally arise due to longitudinal hydrodynamic flow, which establishes a link between spatial and momentum coordinates~\cite{Bjorken:1982qr}. In this regime, both excluded-volume effects and momentum-space interactions become observationally equivalent, since the flow field maps one domain into the other. At lower collision energies, however, the situation becomes considerably more involved; longitudinal boost invariance breaks down, baryon stopping increases, and baryons accumulate in a narrower rapidity interval. As a consequence, the mapping between coordinate and momentum spaces becomes complicated and non-linear. In this regime, the simultaneous success of both excluded-volume and momentum-interaction based models is far from trivial. Therefore, the observed agreement between these conceptually distinct models at low beam energies warrants further detailed theoretical investigation.

\subsection{Baselines from the hadronic transport model UrQMD}
\label{secUrQMD}
Calculations from hadronic transport models are also used as baseline in order to extract genuine properties of the phase structure. Compared to previously discussed thermodynamic calculations, the most obvious advantage of using the transport calculation is that the simulated results can be treated exactly as in experimental data analysis including collision centrality and acceptance for given particles under study. Moreover, essential aspects of the collision dynamics, including volume fluctuations, collective expansion, re-scatterings as well as resonance decays are naturally taken into account.    

In this review, the Ultra-relativistic Quantum Molecular Dynamic (UrQMD) model~\cite{UrQMD} is used as a reference for high moments analysis. The heart of the cascade transport is based on the propagation of hadrons on classical trajectories in combination with stochastic binary scatterings, color string formation, and resonances. The imaginary part of hadron interactions are based on a geometric interpretation of their scattering cross sections~\cite{ParticleDataGroup:2020ssz}. The real part of hadronic interactions is done with a density-dependent potential interaction term based on non-relativistic approaches~\cite{Aichelin:1991xy} with density-dependent Skyrme interactions~\cite{Hartnack:1997ez}. In this model~\cite{Bass:1998ca,Bleicher:1999xi}, for collisions with center-of-mass energies below 5 GeV, particle production is described by interactions between hadrons and resonances, while for $\sqrt{s_{NN}} >$ 5 GeV the excitation of color strings and their fragmentation into hadrons dominates particle production.
 
In the following discussions, the cascade mode of the UrQMD (v3.4) is used. It should be noted that a new version of the UrQMD model (v4.0), in which the chiral mean field (or potential) (CMF) option is available for collisions below $\sqrt{s_{NN}}$ = 5 GeV, has been developed. This extension is important for collectivity of baryons and for the production of light clusters in the higher-density regime~\cite{Bumnedpan:2024jit,Reichert:2022mek,Buyukcizmeci:2023azb,Buyukcizmeci:2023azb,Steinheimer:2025hsr,Steinheimer:2024eha}.

\section{Universality, the QCD crossover, and the critical point}
A central concept in the study of phase transitions in strongly interacting matter is universality, which implies that systems with different microscopic dynamics can exhibit identical critical behavior near a phase transition, provided they share the same global symmetries, dimensionality, and relevant degrees of freedom, see~\cite{Roth:2024hcu} and references there. As a consequence, the nature of the QCD phase transition and its associated critical phenomena can be classified in terms of universality classes, independently of many details of the underlying theory.

In the limit of vanishing up- and down-quark masses, QCD exhibits an exact chiral symmetry, which is spontaneously broken in the vacuum and restored at high temperature~\cite{Koch:1997ei}. In this chiral limit, the finite-temperature transition is expected to be a second-order phase transition belonging to the three-dimensional $O(4)$ universality class, reflecting the symmetry-breaking pattern of two-flavor QCD. This expectation is supported by renormalization-group arguments and lattice-QCD studies. For physical light-quark masses, chiral symmetry is explicitly broken, and the second-order phase transition is replaced by a smooth crossover. Lattice-QCD calculations have established that, at vanishing baryon chemical potential, the transition from hadronic matter to the quark–gluon plasma is not a true phase transition but a crossover occurring over a narrow temperature range~\cite{Borsanyi:2010cj,HotQCD:2014kol}. Despite the absence of singularities in thermodynamic observables, remnants of the underlying $O(4)$ critical behavior persist and manifest themselves in the scaling properties of chiral observables and in higher-order fluctuations~\cite{HotQCD:2018pds}. 

\begin{marginnote}[]
\entry{Crossover, critical point, and universality}
{The QCD crossover at $\mu_{B}\approx 0$ is associated with $O(4)$ universality in the chiral limit. The QCD critical point at finite $\mu_{B}$ is expected to belong to the $Z(2)$ universality class. The liquid--gas critical point is governed by the same $Z(2)$ universality.}
\end{marginnote}

The relevance of universality becomes particularly apparent in the study of fluctuations and correlations of conserved charges. Near a critical point or in the vicinity of a second-order phase transition, fluctuations are governed by long range correlations. As a consequence, the correlation length diverges, leading to characteristic scaling behavior of higher-order cumulants. These scaling properties are universal and determined solely by the universality class of the transition.
In the crossover regime, where the correlation length remains finite, such critical behavior is softened but not entirely erased. Lattice-QCD calculations indicate that higher-order cumulants retain sensitivity to the underlying chiral dynamics and deviate significantly from expectations based on non-interacting or purely hadronic models~\cite{HotQCD:2018pds}. This makes fluctuation observables a powerful tool for probing the remnants of criticality even in the absence of a true phase transition.

At finite baryon chemical potential, the QCD phase diagram is expected to exhibit a critical end point (CEP), at which the crossover transition turns into a first-order phase transition. The existence and location of this critical point remain open questions and are among the primary motivations for experimental programs such as the RHIC Beam Energy Scan.

The QCD critical point is widely believed to belong to the three-dimensional Ising $Z(2)$ universality class, irrespective of the microscopic details of QCD. Near the CEP, thermodynamic observables and fluctuation measures are expected to exhibit characteristic long-range Ising-type critical behavior, including large enhancements of higher-order cumulants and non-monotonic dependencies on the control parameters $T$ and $\mu_{B}$.

The interplay between universality, the QCD crossover, and the critical point has profound implications for heavy-ion experiments. In particular, measurements of higher-order cumulants of conserved charges offer a direct window into critical dynamics and provide a means to discriminate between crossover behavior and proximity to a critical point.

\section{Challenges in experimental measurements}
As mentioned in Section~\ref{int}, the essential idea underlying the connection between experimental measurements and theoretical predictions is motivated by Eq.~\ref{eq_chi}. A key experimental challenge arises from the fact that Eq.~\ref{eq_chi} is formulated for an equilibrated system of fixed volume -  an assumption that is incompatible with realistic experimental conditions, where the volume itself fluctuates from event to event~\cite{Skokov:2012ds, Braun-Munzinger:2016yjz, Rustamov:2022sqm, Holzmann:2024wyd}.

Hence, a meaningful application of Eq.~\ref{eq_chi} requires a rigorous treatment of dynamical variations in the system volume, an issue that becomes increasingly critical for collisions at lower energies~\cite{Braun-Munzinger:2016yjz}.

In addition, due to the smallness of the produced hadron multiplicity used to determine the collision centrality, lower-energy collisions exhibit significantly larger volume fluctuations~\cite{STAR:2022etb}. In contrast, the effect of volume fluctuations becomes small in truly high-energy nuclear collisions~\cite{STAR:2021iop}. The Centrality Bin Width Correction~\cite{Luo:2013bmi} (CBWC) has been widely used in heavy-ion collisions at RHIC energies ($7.7 < \sqrt{s_{NN}} < 200$~GeV) to partially correct measured values of higher-order moments and their ratios for the effects of volume fluctuations. It is important to note, however, that while the averaging procedure employed in the CBWC removes part of the volume fluctuations, the correction is insufficient, particularly for low-energy collisions with $\sqrt{s_{NN}} \lesssim 5$~GeV~\cite{Braun-Munzinger:2016yjz, Friman:2025ulh}.

Moreover, correlations arising from resonance decays may further bias the centrality determination. More detailed discussions of volume fluctuations can be found in Refs.~\cite{STAR:2021iop,STAR:2022etb,Friman:2025ulh,Rustamov:2022sqm,Holzmann:2024wyd}. Recently, a new approach based on an Edgeworth expansion, which determines proton distributions as a function of collision centrality, has been shown to be effective in suppressing volume fluctuations. This method appears to be particularly useful for second- and third-order cumulants and factorial cumulant ratios~\cite{Wang:2025fve} in analyses at fixed-target energies during the second phase of the RHIC Beam Energy Scan.

In this context, care must be exercised when comparing theoretical model predictions to experimental data in cases where volume fluctuations are only partially corrected. In view of these challenges, a prudent strategy for experimental collaborations would be to always publish data uncorrected for volume fluctuations alongside results obtained using specific correction procedures.

\section{Experimental results} 
\label{exp}
\begin{figure}[htb]
\includegraphics[width=6.15in]{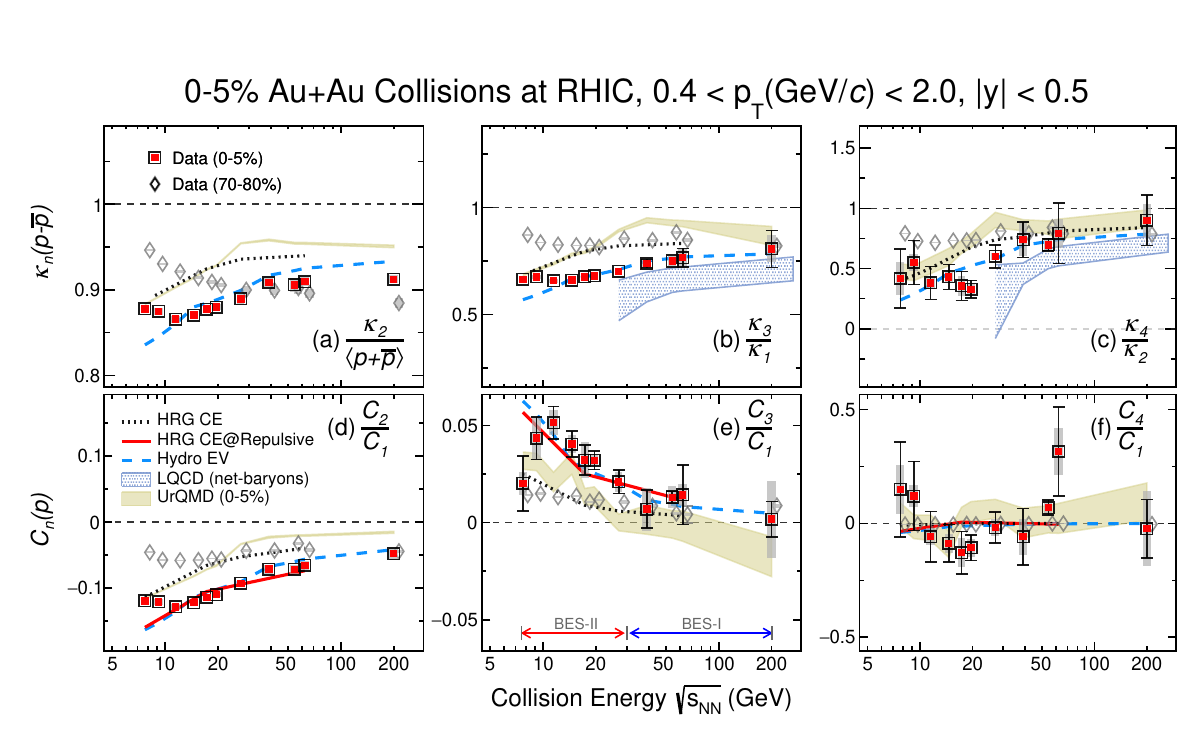}
\caption{Net-proton number cumulant ratios: (a) $\kappa_2/\langle p+\bar{p}\rangle$, (b) $\kappa_3/\kappa_1$, and (c) $\kappa_4/\kappa_2$ and proton number factorial cumulant ratios: (d) $C_2/C_1$, (e) $C_3/C_1$ and (f) $C_4/C_1$ in Au+Au collisions. Results from  BES-II ($\sqrt{s_{NN}}$ = 7.7 -- 27 GeV) and BES-I~\cite{STAR:2020tga,STAR:2021iop} ($\sqrt{s_{NN}}$ = 39 -- 200 GeV) program at RHIC are shown. (Anti-)protons are measured at mid-rapidity ($|y|<0.5$) within the transverse momentum range $0.4 < p_T < 2.0$ GeV/$c$. The bars and bands on the data points reflect statistical and systematic uncertainties, respectively. Theoretical calculations from a hydrodynamical model and repulsive interactions via excluded-volume corrections~\cite{Vovchenko:2021kxx} (Hydro EV, blue dashed line), thermal model with canonical treatment for baryon number~\cite{Braun-Munzinger:2020jbk} (HRG CE, black dashed line), HRG CE with additional repulsive interactions among baryons~\cite{Friman:2025swg}(solid green lines), the UrQMD transport model~\cite{Bass:1998ca,Bleicher:1999xi} ( brown band), and lattice QCD calculations for net-baryon number fluctuations ~\cite{Bazavov:2020bjn,Bollweg:2024epj} (light blue band) are also presented. Figure is taken from~\cite{STAR:2025zdq}.}
\label{fig5}
\end{figure}

Recent STAR results~\cite{STAR:2025zdq} on net-proton number ordinary and factorial cumulant ratios are presented in Fig.~\ref{fig5}, panels (a)--(c) and (d)--(f), respectively. For data recorded in collider mode, identified protons and anti-protons are obtained with high detection efficiency ($\geq 95\%$) and purity ($\sim 99\%$) within mid-rapidity ($|y|<0.5$) and in the transverse momentum range $0.4 < p_T < 2.0~\mathrm{GeV}/c$. Owing to the iTPC upgrade~\cite{ref_iTPC_BESII}, both the pseudo-rapidity coverage and the acceptance at low transverse momentum are significantly extended, resulting in a larger number of charged particles reconstructed in the BES-II program~\cite{STAR:2025zdq}. This enhancement improves the centrality determination. Although the impact on the fourth-order cumulant ratio $\kappa_4/\kappa_2$ remains small due to cancellation effects~\cite{Braun-Munzinger:2016yjz}, the improved acceptance reduces systematic uncertainties across all orders of cumulants and factorial cumulants.

\begin{figure}[htb]
\includegraphics[width=6.15in]{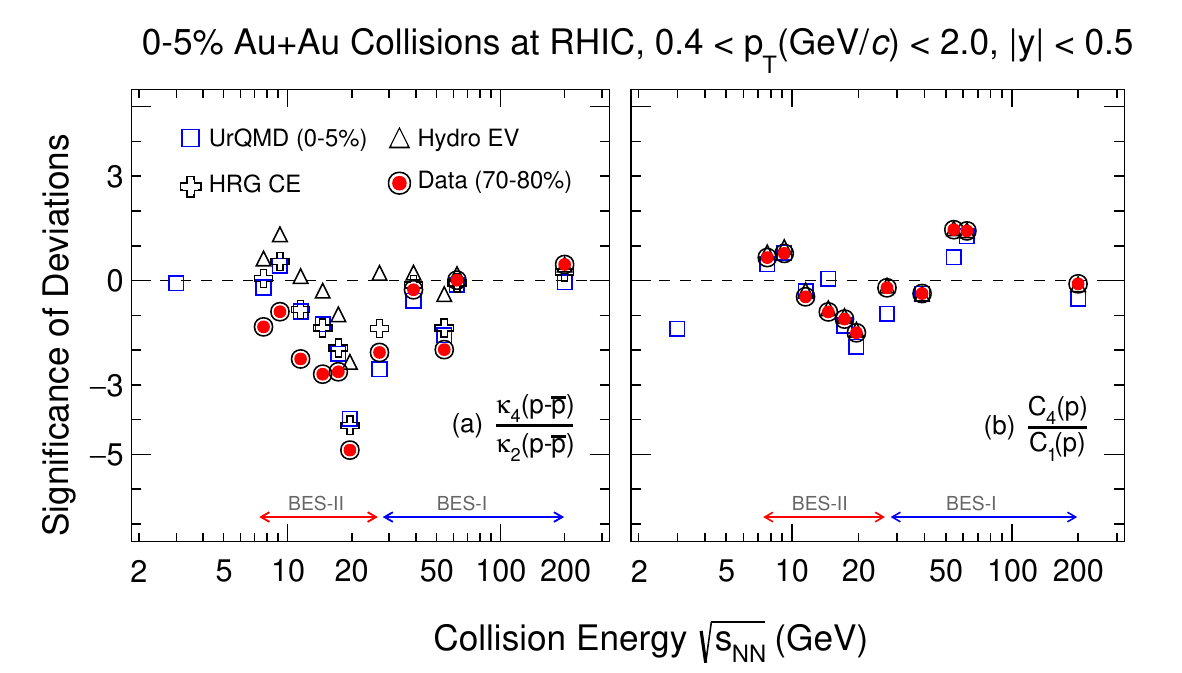}
\caption{Panel(a):Significance of deviation (data$-$reference)/$\sigma_{\rm tot}$ for net-proton cumulant ratios $\kappa_{4}/\kappa_{2}$ in 0-5\% Au+Au collisions~\cite{STAR:2020tga,STAR:2021iop,STAR:2021fge,STAR:2022etb}. References include the non-critical model calculations, such as the UrQMD transport model~\cite{Bass:1998ca,Bleicher:1999xi} (blue squares), HRG with canonical ensemble for baryon number~\cite{Braun-Munzinger:2020jbk} (HRG CE, black crosses), hydrodynamic model with excluded volume~\cite{Vovchenko:2021kxx} (Hydro EV, black triangles), and data from 70-80\% peripheral collisions (red dots). Panel(b): Similar to panel (a) but for factorial cumulant ratio $C_{4}/C_{1}$ for protons. Figure is taken from Ref.~\cite{STAR:2025zdq}.}
\label{fig6}
\end{figure}

Systematic measurements of the collision-energy dependence of ratios of net-proton cumulants and proton number factorial cumulants from BES-II ($7.7 \leq \sqrt{s_{NN}} \leq 27$~GeV) and BES-I ($39 \leq \sqrt{s_{NN}} \leq 200$~GeV) for central (0--5\%) (filled red squares) and peripheral (70--80\%) (open diamonds) collisions are shown in Fig.~\ref{fig5}. 

For comparison, model calculations without critical physics are also shown for the case of central collisions. These include: 

(i) a thermal hadron resonance gas model with canonical ensemble treatment (HRG CE) for baryon number, shown as black-dotted lines~\cite{Braun-Munzinger:2020jbk};
(ii) HRG CE extended to include repulsive interactions among baryons (HRG CE + Repulsive), shown as the solid red lines~\cite{Friman:2025swg};
(iii) a hydrodynamic model with excluded-volume effects (Hydro EV), shown as the dashed blue lines, where the space-time evolution of the medium is simulated assuming local equilibrium with excluded-volume corrections~\cite{Vovchenko:2021kxx}; 
(iv) lattice QCD calculations for net-baryon number cumulant ratios, shown as the blue band, which are compared to the ratios $\kappa_3/\kappa_1$ and $\kappa_4/\kappa_2$ at the high-energy end, $\sqrt{s_{NN}} > 39$~GeV, corresponding to $\mu_B < 200$~MeV~\cite{Bazavov:2020bjn,Bollweg:2024epj}. The overall energy dependence is well captured by the lattice results; however, it should be noted that these calculations are performed for net baryons and do not include experimental acceptance cuts.
(v) the hadronic transport model UrQMD, which is a microscopic approach incorporating non-equilibrium dynamics, hadronic scatterings, and resonance production~\cite{Bass:1998ca,Bleicher:1999xi}. Unlike the thermal and hydrodynamic models, UrQMD does not assume a priori equilibrium, and collision centrality and particle acceptance can be determined consistently with the experimental analysis.

As evident from Fig.~\ref{fig5}, when repulsive interactions among baryons are explicitly included (HRG CE + Repulsive or Hydro EV), the model predictions essentially reproduce the energy dependence of the data from $\sqrt{s_{NN}} = 200$~GeV down to approximately 15~GeV. The implementation of repulsive interactions in both Hydro EV~\cite{Vovchenko:2021kxx} and HRG CE with repulsion~\cite{Braun-Munzinger:2020jbk,Friman:2025swg} is largely phenomenological, and further studies are required to better understand the microscopic origin of these interactions.

Despite the distinct differences among the model approaches, the overall energy dependence of both net-proton number cumulant ratios and proton number factorial cumulants at mid-rapidity for the most central (0--5\%) Au+Au collisions is reasonably well reproduced by non-critical models. At the same time, noticeable discrepancies between the data and these model calculations remain.

To quantify these deviations, we define the significance of the deviation as
\begin{equation}
SD = \frac{\mathrm{data} - \mathrm{ref}}{\sigma_{\mathrm{tot}}},
\end{equation}
where ``ref'' denotes either data from a different collision centrality or a model calculation, and $\sigma_{\mathrm{tot}}$ is the quadratic sum of the uncertainties from the data and the reference. The deviations between the data from central collisions and the references are shown in Fig.~\ref{fig6} as a function of collision energy for the net-proton cumulant ratio $\kappa_4/\kappa_2$ (left panel) and the proton factorial cumulant ratio $C_4/C_1$ (right panel), using data from Ref.~\cite{STAR:2025zdq}.

For the net-proton cumulant ratio $\kappa_4/\kappa_2$, a maximum deviation of about $2$--$5\sigma_{tot}$ is observed at $\sqrt{s_{NN}} = 19.6$~GeV. In Au+Au collisions at $\sqrt{s_{NN}} = 200$~GeV, the system at chemical freeze-out is close to equilibrium~\cite{Andronic:2017pug}, and the fourth-order net-proton cumulant ratio is consistent with the reference calculations. At $\sqrt{s_{NN}} = 3$~GeV, the data are consistent with hadronic transport model calculations~\cite{STAR:2021fge}, supporting the conclusion that hadronic interactions dominate the properties of the medium created in low-energy Au+Au collisions~\cite{STAR:2021fge,STAR:2021yiu,STAR:2021hyx}.

As mentioned in section~\ref{sec_cumulants}, factorial cumulants are sensitive to genuine multi-particle correlations. The deviation of the fourth-order proton factorial cumulant ratio, shown in the right panel of Fig.~\ref{fig6}, also exhibits a minimum at $\sqrt{s_{NN}} = 19.6$~GeV when compared to both peripheral collision data and UrQMD calculations; however, the significance of this deviation is at the level of approximately $2\sigma_{tot}$.

Overall, the fourth-order cumulant ratios show consistency between non-critical thermal model calculations and experimental data at high collision energies, indicating the formation of a thermalized medium at small baryon chemical potential, $\mu_B \sim 0$. In contrast, the most pronounced deviations are concentrated at intermediate collision energies around $\sqrt{s_{NN}} = 19.6$~GeV.

Lower-order ratios of net-proton number cumulants and proton number factorial cumulants also exhibit significant deviations from non-critical references, particularly at lower collision energies corresponding to higher baryon densities. As shown in Fig.~\ref{fig5}, panels (d) and (e), calculations using HRG with repulsive interactions and the Hydro EV model follow the data well from $\sqrt{s_{NN}} = 200$~GeV down to approximately 15~GeV. At lower energies, the model calculations continue to follow the overall trends, while the data become noticeably flatter for $\kappa_2/(p+\bar{p})$ (panel (a)), $\kappa_3/\kappa_1$ (panel (b)), and $C_2/C_1$ (panel (d)). For the third-order proton factorial cumulant ratio $C_3/C_1$ (panel (e)), the data exhibit a more pronounced change around $\sqrt{s_{NN}} = 15$~GeV.

These changes in the energy dependence of the cumulant ratios indicate that repulsive interactions alone are no longer sufficient to describe the medium created in heavy-ion collisions at lower energies. This suggests that the repulsive interaction must either weaken in the low-density regime or that an additional attractive component must be introduced into the model calculations~\cite{Friman:2025swg}.

As discussed in previous sections, evidence from hadron yield measurements (see Figs.~\ref{fig1},~\ref{fig2},~\ref{fig4}), as well as from signatures of collectivity (see Fig.~\ref{fig3}), indicates that a thermally equilibrated medium is formed in high-energy nuclear collisions. Moreover, the energy dependence of the measured proton number cumulants suggests that multiparticle interactions among baryons play a decisive role in shaping the observed fluctuation patterns. At high collision energies, the data are well reproduced by incorporating repulsive interactions, whereas at lower energies attractive interactions may become essential. 

We note that, in the context of a liquid--gas system, the van der Waals equation of state extends the ideal gas law by including short-range repulsion and long-range attraction, thereby naturally giving rise to a first-order phase transition and, more importantly, predicting the existence of a critical point. Further work is needed to elucidate the connection between the liquid–gas transition and the QCD phase transition.

\section{Future Directions} 
\label{future}

Important progress in understanding the QCD phase structure has been achieved at both truly high-energy heavy-ion collisions at the LHC, see discussions in section~\ref{int} and with the RHIC beam energy scan program, see section~\ref{exp}. Small but significant deviations from the non-critical baselines discussed above are seen at center-of-mass collision energies of about 20 GeV and below, lending support to the interpretation that a change in the medium created is observed as the collision energy is lowered. To make further progress, further increase in experimental precision is needed in this energy range, with focus on the high baryo-chemical potential region of $\mu_B \sim$ 600 MeV,
to hunt for possible signs of a 1st order phase transition and the QCD critical point.  

More accurate experiments are also a high priority at $\mu_B \sim$ 0, the LHC region, to map out the properties of the smooth crossover transition. This requires extension of current net-baryon number fluctuations experiments to at least the 6th order for collision systems ranging from pp to central Pb-Pb.

In the somewhat longer term, new experimental facilities are under construction at different accelerator centers:

\begin{enumerate}
\item ALICE 3 at LHC~\cite{Reidt:2024zrb,ALICE:2022wwr} is a recently proposed detector constructed with state-of-the-art pixel technology placed in a superconducting solenoidal magnet. The  innermost two layers of the pixel-tracker will be sitting in a secondary vacuum inside the beam pipe in order to achieve maximum precision for tracking and heavy flavor hadron identification at low transverse momentum in collisions at LHC energy. 

 Important measurements planned with the new device include measurements of quarkonia and heavy-flavor hadrons at low transverse momenta, precision measurements of multi-dimensional dielectron emission to study the evolution of the QGP temperature and multi-charm baryon production to test the dynamics of hadronization and early thermal equilibrium in collisions at LHC energies~\cite{Reidt:2024zrb}.  Important for the subject discussed in this review are very high rate and high precision measurements of net-particle fluctuations near and beyond mid-rapidity.

\item  The FAIR accelerator complex SIS100~\cite{Spiller:2020wtm} offers a maximum proton beam energy of 29 GeV  and very high intensity.  For heavy-ions a projectile kinetic energy range of 4\emph{A} - 11\emph{A} GeV is planned, with 4x$10^{10}$ ions per cycle. The Compressed Baryonic Matter experiment (CBM)~\cite{CBM:2016kpk} at FAIR is a next generation high rate fixed-target experiment capable of handling interaction rates  up to 10 MHz. In this experiment, the target is placed at the entrance of a superconducting dipole magnet and the downstream tracking system is made of a micro-vertex detector, followed by a silicon tracking system that serves as the main tracking detector, both placed inside the magnet. Directly after the magnet are RICH and TRD detectors for election identification. For muon identification, the RICH is replaced by hadron absorbers with gas tracking detectors (MuCH). Charged hadron identification in the CBM experiment is achieved by combining high-precision momentum measurements, specific energy-loss information from the Silicon Tracking System, and particle velocity. The latter is determined from the reconstructed track length together with the time-of-flight measurement provided by the Time-of-Flight system, which is based on multi-gap resistive plate chambers. A forward spectator detector (FSD) will provide crucial information on collision centrality and event-plane. 

As discussed in Ref.~\cite{CBM:2016kpk}, the main focus of the CBM experiments are dileptons, both dielectrons and dimuons, strangeness and, for pA collisions, charm physics. Baryon correlations and fluctuations are also important at the high baryon density region. The CBM experiment will cover the center-of-mass energy per nucleon pair in the range of 2.5 - 4.9 GeV, for Au+Au collisions, which is potentially crucial for the search of the QCD critical point. Most recently, theoretical speculations~\cite{Clarke:2024ugt, Basar:2023nkp, Hippert:2023bel, Fu:2019hdw, Gunkel:2021oya, Gao:2020fbl} as well as data driven analysis~\cite{Sorensen:2024mry} are all focused on an energy region where  the QCD critical point ($T \sim 100\pm 20$ MeV, $\mu_B \sim 600\pm50$ MeV) could be situated.
\end{enumerate}

Finally we would like to point out that, while RHIC covers the finite region ($160 > T >100$ MeV, $25 < \mu_B < 750$ MeV) in the QCD phase diagram, ALICE 3 at the LHC and CBM at FAIR cover two extreme regions ($T\sim 160$ MeV, $\mu_B\sim0$ MeV) and ($100 > T > 65$ MeV, $550 < \mu_B < 800$ MeV), respectively. Although these regions appear widely separated in the phase diagram, a full understanding of the QCD phase diagram requires the study of both regions, as they are thermodinamically connected. With the expertise acquired with the RHIC and FAIR experiments at low energies and with ALICE at the LHC we look forward to decisive new insights from these next generation experiments on the phase structure of QCD.

\section*{DISCLOSURE STATEMENT}
The authors are not aware of any affiliations, memberships, funding, or financial holdings that
might be perceived as affecting the objectivity of this review. 

\section*{ACKNOWLEDGMENTS}
We thank Drs. X. Dong, K. Redlich, and J. Stachel for insightful discussions.

\printbibliography
\end{document}